\documentclass{iopart}

\pdfoutput=1

\usepackage[english]{babel}
\usepackage[utf8]{inputenc}
\usepackage[colorlinks=true,linkcolor=blue, citecolor=blue, urlcolor=blue, bookmarks]{hyperref}
\usepackage{slashed}
\usepackage{amssymb}
\usepackage{tikz}

\usepackage{cite}

\def\be{\begin{equation}}
\def\ee{\end{equation}}

\def\bea{\begin{eqnarray}}
\def\eea{\end{eqnarray}}

\def\Tr{{\rm Tr}}
\def\iu{{\rm i}}

\begin{document}

\title[Entanglement spectrum degeneracy and Cardy formula in CFT]{Entanglement spectrum degeneracy and Cardy formula in $1+1$ dimensional conformal field theories}

\author{Vincenzo Alba, Pasquale Calabrese, Erik Tonni}

\address{ SISSA and INFN, Via Bonomea 265, 34136 Trieste, Italy 
}

\begin{abstract}

We investigate the effect of a global degeneracy in the distribution of entanglement spectrum in conformal field theories  in one spatial dimension. We relate the  recently found universal expression for the entanglement hamiltonian to the distribution of the entanglement spectrum. The main tool to establish this connection is the Cardy formula. It turns out that  the Affleck-Ludwig non-integer degeneracy, appearing because of the boundary conditions induced at the entangling surface, can be directly read from the entanglement spectrum distribution. We also clarify the effect of the non-integer degeneracy on the spectrum of the partial transpose, which is the central object for quantifying the entanglement in mixed states. We show that the exact knowledge of the entanglement spectrum in some integrable spin-chains provides strong analytical evidences corroborating our results.  

\end{abstract}

\maketitle

\section{Introduction}

In the course of the noughties, entanglement has become a standard and powerful 
tool for the study of many-body quantum systems, both in and out of thermodynamic 
equilibrium (see, e.g., Refs. \cite{amico-2008,calabrese-2009,eisert-2010,laflorencie-2016} 
as reviews); for example, entanglement is nowadays routinely used for the identification 
of critical and topological phases of matter.
The central quantity to quantify the bipartite entanglement in a many-body system is 
the reduced density matrix $\rho_A$ of a subsystem $A$, which is defined as 
$\rho_A={\rm Tr }_B \rho$, where $B$ is the complement of $A$, and $\rho$ 
is the density matrix of the entire system. When the system is in a pure state 
$|\Psi\rangle$ and $\rho=|\Psi\rangle\langle\Psi|$, the entanglement can be measured 
by the von Neumann or R\'enyi entropies of $\rho_A$.
However, it has been pointed out long ago by Li and Haldane~\cite{lh-08} that the 
reduced density matrix (RDM) encodes much more information than the entanglement entropy. 
Part of this information may be extracted by looking at the entire spectrum of $\rho_A$, 
which has been dubbed {\it entanglement spectrum}. 
This observation triggered a systematic study of the entanglement spectrum, which proved 
to be an extremely powerful theoretical tool to analyse topological phases~\cite{lh-08,
topo1,topo2,topo4,bulkedge,herm-17}, symmetry-broken phases~\cite{metlitski-2010,alba-2013a,
kolley-2013,kolley-2015,frerot-2016}, disordered systems~\cite{fcm-11,mbl-es,mbl-2,ppmps-17}, 
and gapless one-dimensional phases~\cite{cl-08,lauchli-2013,lbl-16}. Furthermore, a 
protocol for measuring the entanglement spectrum in cold-atom experiments has been 
proposed~\cite{pzszh-16}, which generalises the recent measurements of entanglement 
entropy~\cite{islam,kauf}.

In Ref. \cite{cl-08}, it has been pointed out that the distribution of the eigenvalues $\lambda_i$
of the reduced density matrix $\rho_A$ (also known as entanglement spectrum distribution)
\be
P(\lambda)\equiv \sum_i \delta(\lambda-\lambda_i),
\ee
can be reconstructed from the analytic knowledge of the moments ${\rm Tr} \rho_A^n$ of $\rho_A$. 
In this respect, conformal invariant field theories represent a very useful playground. 
Indeed, for a conformal invariant  system in one spatial dimension, for a finite interval $A$ of length $\ell$ 
embedded in an infinite system, the moments  are given as~\cite{cc-04,cc-09}
\be
{\rm Tr} \rho_A^n=c_n \Big( \frac{\ell}{\epsilon} \Big)^{c/6(n-1/n)},
\label{mom-cft}
\ee
where $c$ is the central charge of the conformal field theory (CFT), $\epsilon$ an 
ultraviolet cutoff, and $c_n$ a {\it non-universal} and in general unknown 
$n$-dependent amplitude. 
Assuming  $c_n$ to be constant, in Ref.~\cite{cl-08} a super-universal form of the 
entanglement spectrum distribution has been explicitly worked out: it turned out to 
depend only on the central charge via the largest eigenvalue of $\rho_A$. Thus, 
when expressed in term of the latter, the entanglement spectrum distribution does not depend on 
any parameter of the theory.
This super-universal distribution leads to an extremely simple scaling law for the 
integrated distribution function $n(\lambda)$, i.e. the number of eigenvalues 
larger than $\lambda$, which is 
\be
n(\lambda)= I_0 (\xi_\lambda)\,, \qquad \xi_\lambda \equiv 2\sqrt{(-\ln \lambda_{\rm max})
\ln(\lambda_{\rm max}/\lambda)}\,,
\label{n-lam}
\ee
where $I_0(x)$ is the modified Bessel function and $\lambda_{\rm max}$ the 
largest eigenvalues of $\rho_A$. 
Obviously, the only condition for the validity of (\ref{n-lam}) is the particular $n$-dependence in~(\ref{mom-cft}). 
Consequently, the same law (\ref{n-lam}) is valid in many other situations in CFT (finite systems, systems with boundaries, 
finite temperature, etc), 
as well as in proximity of quantum critical points, where the same scaling (\ref{mom-cft}) holds with the replacement of $\ell$ with the 
correlation length \cite{cc-04}.
Several numerical studies~\cite{cl-08,pm-10,ahl-12,lr-14,si-14,laflorencie-2016} 
tested that the super-universal distribution~\eref{n-lam} describes surprisingly 
well the spectrum of lattice models whose low-energy spectrum has an underlying CFT, 
despite  the assumption of $c_n$ being constant. The exact knowledge of the 
distribution (\ref{n-lam}) played also a central role in the understanding of the 
performance of matrix product states algorithms~\cite{chisc}. 

In very recent times, a lot of activity has been devoted also to the study of the 
operatorial form of the reduced density matrix~\cite{ncc-08,ch-09,pc-11,chm-11,
ls-12,wkpv-13,hshf-14,lla-14,p-14,ct-16,kvw-15,ep-17,kvw-17,dvz-17}. To this goal, the 
reduced density matrix is written as  
\be
\rho_A = \, e^{-2\pi K_A},
\ee
where $K_A$ is the entanglement (or modular) hamiltonian. 
The first study of the 
modular Hamiltonian dates back to the seminal work by  Bisognano and Wichmann~\cite{bw-76}, 
which proved that for an arbitrary relativistic quantum field theory in Minkowski space of generic dimensionality
and for a bipartition in two equal semi-infinite parts separated by an infinite hyperplane,
the entanglement Hamiltonian may be expressed as an integral of the local energy 
density with a space-dependent weight factor.
This theorem has been  exploited, 
especially in  CFTs in  arbitrary dimension, to relate entanglement Hamiltonians 
in different bipartitions~\cite{chm-11, wkpv-13, ct-16}. In turn, these works relate the 
entanglement to the Hamiltonian spectrum, a result which we will exploit here.

This paper has many different goals which can be summarised as follows. 
On the one hand, we want to directly derive the 
entanglement spectrum distribution in~\cite{cl-08} from the entanglement Hamiltonian 
in~\cite{ct-16}, also to understand the validity of the assumptions made in both~\cite{ct-16} 
and~\cite{cl-08}; in particular, the assumption of $c_n$ being constant (see~\eref{mom-cft}). 
We anticipate that the relation between the  distribution of the entanglement spectrum 
and that of the energy spectrum of the CFT is the famous Cardy formula~\cite{c-86}. 
This naturally poses the question about the effect of degeneracies of the spectrum of $\rho_A$, 
which must be related to the Affleck-Ludwig boundary entropy~\cite{al-91} appearing 
in the Cardy formula on the annulus. In turn, this leads to a non-trivial $n$-dependence 
of the factor $c_n$ in (\ref{mom-cft}), which affects the entanglement spectrum distribution.
Thus this result shows that the physics at the entangling surface can be read off from the entanglement spectrum distribution.
Finally,  we will show that the degeneracy in the spectrum of $\rho_A$ 
has non-trivial effects also on the distribution of the eigenvalues of the partial transpose
of $\rho_A$, i.e., the negativity spectrum of Refs.~\cite{rac-16,mac-17}.

This paper is organised as follows. In Sec. \ref{sec2}, as a warming up exercise, we work 
out with elementary methods the consequences of a global degeneracy in the entanglement 
spectrum on the distribution of eigenvalues and on the moments of the reduced density 
matrix. In Sec. \ref{sec3} we show that the entanglement spectrum distribution in~\cite{cl-08} 
is a reparametrisation of the Cardy formula. This is one of the main result of this paper 
that allows us to understand also how the Affleck-Ludwig boundary entropy~\cite{al-91}, induced 
by the physics at the entangling surface, affects the entanglement spectrum distribution. 
In Sec. \ref{sec4}, we test the general results of the previous sections in some integrable spin-chains. 
In Sec. \ref{sec5} we explore the consequences of the degeneracy of the entanglement spectrum for the 
spectrum of the partially transposed density matrix, i.e., for the negativity spectrum. 
Finally in Sec. \ref{sec6} we draw our conclusions.

\section{Entanglement spectrum distribution and global degeneracies}
\label{sec2}

We start by considering an elementary exercise about the effects of global degeneracy of 
the eigenvalues of the reduced density matrix (e.g., as a consequence of some internal 
or topological symmetry) on the entanglement spectrum distribution. This problem can be 
understood with several equivalents methods, but in the following we prefer to exploit 
the relation of the entanglement spectrum with the moments of the reduced density 
matrices (i.e., the R\'enyi entanglement entropy) so to make a direct contact with 
the derivation of the CFT entanglement spectrum distribution in Ref.~\cite{cl-08}. We stress that 
there are no new findings in this section, but just a different view about well known facts. 

Let us consider the case in which all eigenvalues $\lambda_j$  of the RDM have a degeneracy $g$.
Here $j=0,1,2,\dots ,Ng-1$, and the $g$ consecutive eigenvalues with 
$ gi \leqslant j  < g(i+1)$ with $i=0,1,2\dots ,N-1$  are all equal. 
We also introduce a set of non-degenerate eigenvalues 
$\mu_i$ (with $i=0,1,2\dots N-1$), just taking one every $g$-degenerate eigenvalues $\lambda_j$. 
Because of the normalisation $\Tr \rho_A=1$, the only way of doing so is by rescaling the 
eigenvalues by a factor $g$, i.e., 
\be
\mu_i=g \lambda_{gi},  
\ee
so that
\be
\sum_i \mu_i=1\Longleftrightarrow \sum_j \lambda_j = \sum_i g \frac{\mu_i}g =1\,.
\ee
For the remaining part of this section, we will denote the RDM with eigenvalues $\lambda_j$ as 
$\rho_\lambda$ and the other with eigenvalues $\mu_i$ as $\rho_\mu$.  

Clearly, the  following relation for the moments of the reduced density matrix holds 
\bea\fl\nonumber
R_n(\mu) \equiv \Tr \rho_\mu^n= \sum_i \mu_i^n \\
\Rightarrow R_n(\lambda)\equiv \Tr \rho_\lambda^n= \sum_j \lambda_j^n= \sum_i g 
\Big( \frac{\mu_i}g\Big)^n= g^{1-n} R_n(\mu)\,.
\label{R-rel}
\eea
For conformal field theories, this degeneracy does not alter the large $\ell$ dependence of the 
moments $R_\mu$ (see \eref{mom-cft}), but it provides a relation between the multiplicative constants 
$c_n$ in \eref{mom-cft}, such that 
\be
c_n^{(\lambda)}= g^{1-n} c_n^{(\mu)}.
\label{c-lam-mu}
\ee
Eq.~\eref{c-lam-mu} is valid whenever a scaling like \eref{mom-cft} occurs, 
e.g. close to a conformally invariant quantum critical point~\cite{cc-04}. 
For the R\'enyi entropy $S_n\equiv \frac1{1-n} \ln \Tr \rho_A^n$, Eq.~(\ref{R-rel}) implies 
\be
S_n(\lambda)= S_n(\mu)+ \ln g\,,
\label{Splusg}
\ee
i.e., the degeneracy gives an additive constant which does not depend neither on $\ell$ 
nor on $n$.
This indeed is the well-known {\it topological term} due to the degeneracy of the 
entanglement spectrum~\cite{top-ent}.  

We now explore the consequences of this degeneracy on the entanglement spectrum 
distribution $P_\lambda(\lambda)=\sum \delta(\lambda-\lambda_i)$ that can be reconstructed 
from the knowledge of the moments $R_n(\mu)$~\cite{cl-08}. Indeed, after introducing the 
Stieltjes transform of $\mu P_\mu(\mu)$ 
\begin{equation}
\label{fz}
f_\mu(z)\equiv\frac{1}{\pi}\sum_{n=1}^\infty R_n(\mu) z^{-n}=\frac{1}{\pi}\int d\mu 
\frac{\mu P_\mu(\mu)}{z-\mu}, 
\end{equation}
one has 
\be
\mu P_\mu(\mu)=\lim_{\epsilon\to 0}\textrm{Im} f(\mu-i\epsilon).
\ee
The relation between $P_\mu(\mu)$ and $P_\lambda(\lambda)$  easily  follows from
\be
f_\lambda(z)\equiv\frac{1}{\pi}\sum_{n=1}^\infty R_n(\lambda) z^{-n}=
\frac{1}{\pi}\sum_{n=1}^\infty g^{1-n} R_n(\mu) z^{-n}= g f_\mu (zg).
\ee
For the eigenvalue distribution this implies 
\be\fl
\frac{1}{\pi}\int d\lambda \frac{\lambda P_\lambda(\lambda)}{z-\lambda} = f_\lambda(z)=
g f_\mu (zg)= \frac{g}{\pi}\int d\mu \frac{\mu P_\mu(\mu)}{zg-\mu}= 
 \frac{1}{\pi}\int d\lambda g^2 \frac{\lambda P_\mu(\lambda g)}{z-\lambda},
\ee
i.e., the final relation is
\be
P_\lambda(\lambda) = g^2 P_\mu (g \lambda). 
\label{Plm}
\ee
Notice that $\int d\lambda \lambda P_\lambda(\lambda)=1$ {follows} as a consequence of the normalisation of $P_\mu(\mu)$. 
The factor $g^2$ (instead of $g$) at first can look awkward, but it is clearly due to the Jacobian necessary also 
for the normalisation.  

As already mentioned in the introduction, in CFT a central quantity is the 
number distribution function of the entanglement spectrum, which is defined as 
the number of eigenvalues of the RDM larger than $\lambda$, i.e.,  
\be
n_\lambda(\lambda)\equiv \sum_{\lambda_i> \lambda} 1= \int_{\lambda}^{\lambda_{
\rm max}} P_\lambda(x) d x\,. 
\ee
The relation between $n_\mu(\mu)$ and $n_\lambda(\lambda)$ is 
\be\fl
n_\lambda(\lambda)= \int_{\lambda}^{\lambda_{\rm max}} P_\lambda (x)d x=
g^2 \int_{\lambda}^{\lambda_{\rm max}} P_\mu(g x) d x= 
g \int_{\lambda g}^{\mu_{\rm max}}  P_\mu(y) d y= g n_{\mu} (\lambda g),
\ee 
where we used $\lambda_{\rm max}=\mu_{\rm max}/g$.

In a CFT with the assumption $c_n=1$ (implying a non-degenerate spectrum), 
it has been shown \cite{cl-08} that  $n_{\mu}(\mu)$ can be written as (\ref{n-lam}).
Then, for a model with a low-energy spectrum described by a CFT, but with a 
global degeneracy $g$, since $ (\mu_{\rm max}/\mu) =  (\lambda_{\rm max}/\lambda)$, 
at the leading order in $\ln \ell$ for large $\ell $, we have
\be
n_\lambda(\lambda)\simeq g I_0(\xi_\lambda),
\label{g-n}
\ee
i.e., in the rescaled variables, the eigenvalue distribution gets multiplied 
by the degeneracy. This is a quite trivial result that we found instructive to 
derive within the methods of Ref.~\cite{cl-08}. 
Since $I_0(z)$ for large positive $z$ grows exponentially, 
usually it is convenient to plot the logarithm of the number distribution, so 
that the net effect of $g$ is an additive constant $\ln g$. 

\section{Entanglement spectrum distribution from the Cardy formula}
\label{sec3}

\begin{figure}[t]
\begin{center}
\includegraphics*[width=0.65\linewidth]{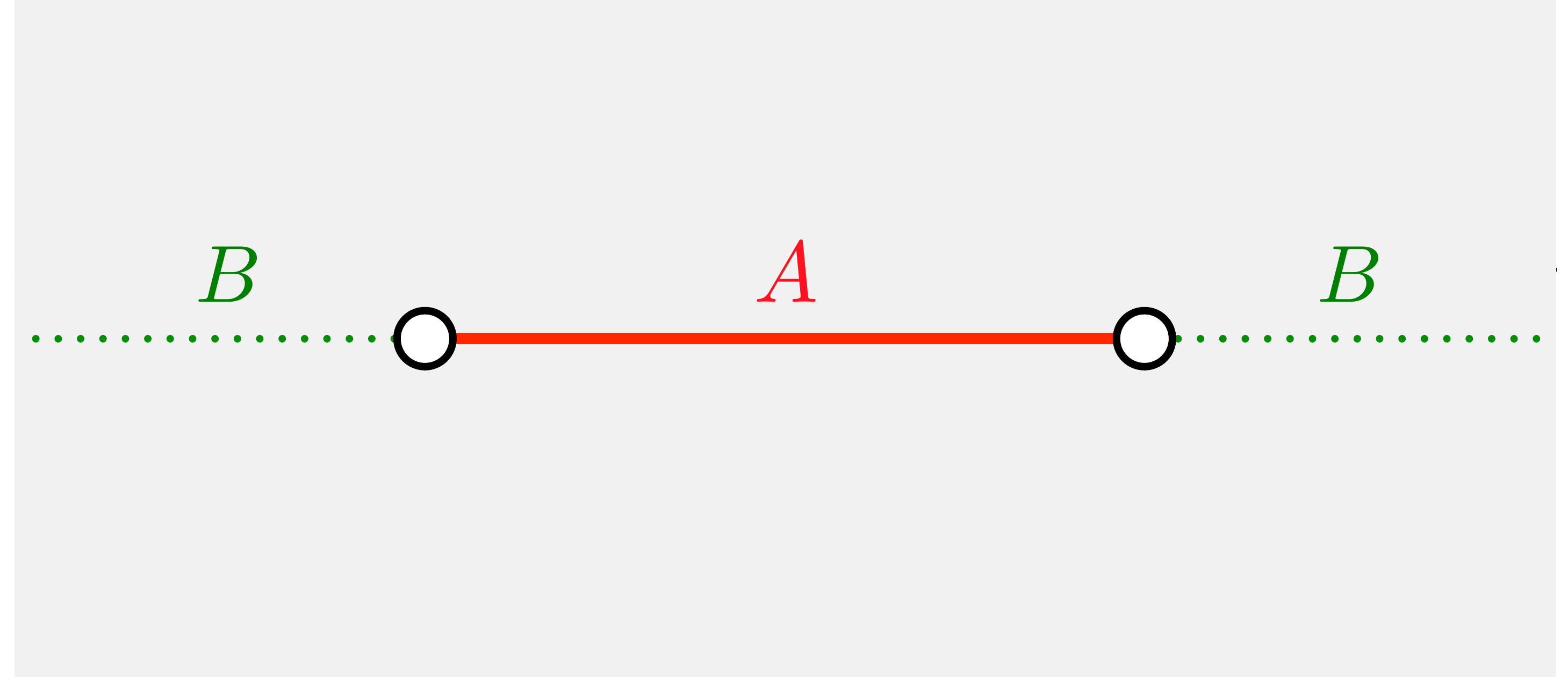}
\end{center}
\caption{ Path integral for reduced density matrix $\rho_A$ of an interval of 
 length $\ell$ embedded in an infinite system: The rows and columns 
 of the density matrix are labelled by the values of the fields on 
 the upper and lower edges of the slit along $A$. The field theory 
 expectation value is made finite by eliminating two circular regions 
 of radius $\epsilon$ at the two boundary points (entangling surface) 
 with conformally invariant boundary conditions. The moments ${\rm Tr}
 \rho_A^n$ are computed by joining cyclically, along $A$,  $n$ 
 replicas of $\rho_A$, thus leaving small holes at the entangling 
 surface. The resulting manifold has the  topology of an annulus.
}
\label{fig1}
\end{figure}

In this section we show how the entanglement spectrum distribution in a $1+1$ 
dimensional conformal field theory~\cite{cl-08} can be obtained from 
the CFT density of states.
In fact, we will show that the entanglement spectrum distribution is just a 
reparametrisation of the former. The main tool to obtain this result is 
the  universal form for the CFT entanglement hamiltonian which has been recently found in \cite{ct-16}.
The leading term of the density of states for asymptotically large energies is the  famous Cardy formula~\cite{c-86}, 
which during the years has been investigated by many authors in order to include subleading terms~\cite{al-91,carlip-98,carlip-00,loran}.

As explained in the introduction, the reduced density matrix can be always written as  
\be
\rho_A = \, e^{-2\pi K_A},
\ee
where $K_A$ is the entanglement hamiltonian. 
For a 1+1 dimensional CFTs, in the case when $A$ is a finite interval of length $\ell$ 
embedded in an infinite system, $\rho_A$ is given by a path integral 
which is pictorially reproduced in Fig.~\ref{fig1}.
The rows and columns of the density matrix are labelled by the values of the fields on the upper and lower edges of the slit along $A$,
while along $B$ the field is continuous. 
In any field theory, this representation of the reduced density matrix is plagued by ultraviolet divergences 
originated by the degrees of freedom in the vicinity of the boundary points between $A$ and $B$
(generally the boundary between $A$ and $B$ is called {\it entangling surface} and we will use this terminology, 
although in the case of interest here the surface is just made of one or  two  points). 
In a CFT, a very convenient way of regularising these ultraviolet divergencies is to remove 
small disks of radius $\epsilon$ (the ultraviolet cutoff) around the endpoints of the interval~\cite{hlw-94,ot-15} (see Fig.~\ref{fig1}).
The price to pay is that some {\it conformal invariant} boundary conditions must be imposed on these disks. 
At this point,  the spacetime in Fig. \ref{fig1} has the topology of an annulus.
Indeed, the mapping between this spacetime and the annulus can be explicitly worked out and 
the resulting width $W$ of the annulus is related to $\ell$ as $W=2\ln(\ell/\epsilon) + O(\epsilon)$~\cite{hlw-94,ct-16}. 
An unexpected and surprising result is that the boundary conditions imposed at the small disks affect the value of 
physical quantities. In particular the R\'enyi entanglement entropies are \cite{ct-16}
\be
S_n= \frac{c}6\left(1+\frac1n\right)\ln \frac\ell\epsilon+ \ln g,
\label{Svsg}
\ee
where the first term is the well known universal leading logarithmic term \cite{hlw-94,cc-04}
and $\ln g$ is the Affleck-Ludwig boundary entropy corresponding to the boundary conditions on the small disks. 
This should be regarded as surprising because the ultraviolet regularisation of the theory appears to affect physical observables. 
The explanation for this fact is that there is some physics at the entangling surface (see also \cite{ot-15}):  
the idea we have in mind is that in a given microscopical model whose low-energy physics is described by a CFT,
the partial trace over $B$ induces boundary conditions (which could depend on the actual degeneracy of
the ground-state of the model that could not be unique as in a bulk CFT) and these affect physical quantities as in~\eref{Svsg}.
We stress that the $\ln g$ terms \eref{Svsg} is not  the same as the one found in \cite{cc-04} which was instead 
due to the boundary entropy of the physical boundary of a CFT. 
This is instead analogous to Eq. \eref{Splusg} where $\ln g$ enters as a true degeneration of the entanglement spectrum. 
The main difference between the two is that in CFT the degeneration $g$ can be non integer. 

At this point one could naively think that the actual value of $\ln g$ in a given microscopical model can be always read out from the 
analysis of the R\'enyi entropies. This unfortunately is not the case because generically the additive constant term of the 
entanglement entropy gets non-universal contributions from ultraviolet physics in the bulk which are very difficult to disentangle from 
the $\ln g$ term. This is evident if one tries to extract $\ln g$ from some analytically known cases, as e.g. those in 
Refs. \cite{jk-04,ij-08,ce-10,cmv-11,dsvc17} and becomes even more cumbersome in numerics. 
(Anyhow, there are some cases when instead the appearance of $\ln g$ is clear and these 
will be considered in the next sections to substantiate some of our findings in this section.)
One of the consequence of the following analysis is that the actual value of $g$ should be accessible in a easier manner 
from the analytic or numerical study of the entanglement spectrum distribution.

After this long discussion,  we are ready to relate the CFT density of states and the entanglement spectrum distribution.
In Ref.~\cite{ct-16}  the entanglement hamiltonian has been related to the generator of the translations around an annulus 
with the same conformally invariant boundary condition imposed on the boundaries of the small disks.
By exploiting  known results on the annulus, the eigenvalues $\kappa_\Delta$ of $K_A$ have been written as~\cite{ct-16}
\be
\label{kappa}
\kappa_\Delta = \frac{\pi}{W} \left( \Delta - \frac{c}{24} \right) + \textrm{const},
\ee 
where $\Delta$ are the conformal dimensions of the operators of the CFT 
(both the primaries and their descendants). 
Eq.~\eref{kappa} implies that the eigenvalues $\lambda_\Delta $ of $\rho_A$ are in one to one 
correspondence with the spectrum of operators of the CFT  as
\be
\label{lambda_Delta}
\lambda_\Delta 
= \, e^{- 2\pi \kappa_\Delta}
=\, \lambda_0 \, e^{- 2\pi^2 (\Delta - c/24)/W},
\ee
where the constant in (\ref{kappa}) has been absorbed in $\lambda_0$. To fix the constant 
$\lambda_0$, let us consider the maximum eigenvalue $\lambda_{\rm max}$ of the RDM, 
which is obtained for the minimum value of $\Delta$ denoted by 
$\Delta_{\rm min}$:
\be\fl
\label{lambda_max}
\lambda_{\Delta_{\rm min} }
=\, \lambda_0 \, e^{- 2\pi^2 (\Delta_{\rm min} - c/24)/W}
\;\stackrel{W \gg 1 }{\longrightarrow}\;\lambda_0
\qquad \Longrightarrow \qquad
\lambda_0 = \lambda_{\rm max} ,
\ee
i.e., the normalisation constant $\lambda_0$, in the limit of large $\ell$, is nothing 
but the largest eigenvalue of the RDM.

In the limit of large but finite $\ell$ (or equivalently, large $W$), 
the eigenvalues of the RDM form a continuum and their asymptotic distribution has been determined analytically in~\cite{cl-08}. 
In this manuscript we reobtain this result by using the Cardy formula~\cite{c-86}. 
Our analysis allows to find that the entanglement spectrum distribution depends also on the Affleck-Ludwig ground state degeneracy 
 \cite{al-91} which originates from the conformally invariant boundary states characterising the CFT on the annulus. 
 This important feature has been overlooked in all the literature about the entanglement spectrum.

As first step, let us write down explicitly the inverse function $\Delta = \Delta(\lambda)$. From (\ref{lambda_Delta}) we find
\be
\label{delta_sub}
\Delta - \frac{c}{24} 
\,=\,\frac{W}{2\pi^2}\, \ln(\lambda_{\rm max}  / \lambda_\Delta )
\,=\,\frac{6 b}{\pi^2 c}\, \ln(\lambda_{\rm max}  / \lambda_\Delta ),
\ee
where, as in Ref.~\cite{cl-08}, we have introduced the parameter $b$ as $b\equiv 
{c W}/{12}=-\ln\lambda_{\rm max}$. Actually, since we are considering the limit 
of large $\Delta$, the factor $c/24$ is a subleading correction, which we keep 
since it does not influence the final result. 

Given the correspondence (\ref{lambda_Delta}) between the eigenvalues of the 
RDM and the conformal spectrum, the distribution of eigenvalues $P(\lambda)$ of 
the RDM is 
\bea\fl 
\label{distribution P0}
P(\lambda) 
&=&
\int_{\Delta_{\rm min}}^{\infty}
\rho(\Delta)\, \delta(\lambda - \lambda_\Delta)\, d\Delta
\;=\;
\int_{\lambda_{\rm max}}^{0}
\rho(\Delta)\, \delta(\lambda - \lambda_\Delta)\,  \frac{d \Delta }{d \lambda_\Delta} \, d\lambda_\Delta
\nonumber\\\fl
& =&
- \theta(\lambda_{\rm max} - \lambda)\, 
\frac{d \Delta(\lambda) }{d \lambda}
\, \rho(\Delta (\lambda)),
\label{distribution P}
\eea
where  $\rho(\Delta)$ is the density of states in the CFT. 
This was  introduced by Cardy~\cite{c-86} who derived for the first time its large $\Delta$ behaviour for a bulk CFT, known 
nowadays as  the Cardy formula.

In the \ref{sec:app} we have employed the modular invariance on the annulus to obtain a suitable expression of $\rho(\Delta)$ 
for large $\Delta$. 
In order to have a straightforward comparison with \cite{cl-08}, we have to take into account also higher order terms in the expansion
(see \cite{loran} for a similar calculation on the torus).
The result reads (see (\ref{inv-lap}))
\be \fl
\label{rho}
\rho(\Delta) 
\simeq 
g \, \frac{\pi^2 c}{3}\; \frac{I_1(\sigma(\Delta))}{\sigma(\Delta)} + g \delta(c-\Delta/24),
\qquad {\rm with} \quad
\sigma(\Delta)
\,\equiv\,
2\pi  \sqrt{\frac{c}{6}\left(\Delta - \frac{c}{24} \right)} ,
\ee
where $I_1$ is the modified Bessel function of the first kind, $c$ is the central charge of the underlying CFT and
$g\equiv g_a g_b$ is the ground-state degeneracy induced by the 
non-trivial boundary conditions (which in a CFT can be non-integer). 
Here, $g_a\equiv \langle 0|a\rangle$ and $g_b\equiv \langle 0|b\rangle$ are 
the degeneracies introduced by the two boundaries of the annulus.
In the case of interest $g_a = g_b$.
Exponentially suppressed contributions have been neglected in (\ref{rho}).
By expanding (\ref{rho}) for large $\Delta$ and taking the first subleasing correction after the leading exponential term, the result of \cite{al-91}
is recovered (see (\ref{rho AL order1})).

Eq.\,(\ref{delta_sub}) allows to write the function $\sigma(\Delta)$ in (\ref{rho}) as function of 
$\lambda$, obtaining 
\be
\sigma(\lambda)=
\sigma(\Delta(\lambda))=2 \sqrt{b \ln(\lambda_{\rm max}  / \lambda) }\,,
\ee
with $b=-\ln \lambda_{\rm max}$.
Finally, plugging the Cardy formula (\ref{rho}) for the density of states, in the 
distribution of the eigenvalues of the RDM (\ref{distribution P}) we have
\bea
\label{distribution P ris}
P(\lambda) \,
&=&\,
\theta(\lambda_{\rm max} - \lambda)\,
\frac{6 b}{\pi^2 c \,\lambda}
 \left[\, g\, \frac{\pi^2 c}{3} \; \frac{I_1(\sigma( \lambda))}{\sigma(\lambda)} \,\right]+ g \delta(\lambda-\lambda_{\rm max})
 \nonumber
\\
&=&\,
 \theta(\lambda_{\rm max} - \lambda)  \;  g\,\frac{b}{ \lambda }\,
\frac{I_1\big( 2 \sqrt{b \ln(\lambda_{\rm max}   / \lambda ) }  \big)}{
\sqrt{b \ln(\lambda_{\rm max}   / \lambda) }}+ g \delta(\lambda-\lambda_{\rm max})\,,
\eea
which is  exactly the same as the entanglement spectrum distribution of \cite{cl-08} 
multiplied by $g$. 
 Denoting by $P_\mu(\mu)$ the above distribution for $g=1$ and with $P_\lambda(\lambda)$ the one for arbitrary $g$,
we have that Eq. (\ref{distribution P ris}) satisfies the relation (\ref{Plm}) --found for integer $g$-- in the limit $-\ln\lambda_{\rm max}\gg1$.
Eq. (\ref{distribution P ris}) generalises then the result of the previous section to 
the Affleck and Ludwig non integer degeneracy  \cite{al-91}, showing that it
influences the entanglement spectrum distribution in a sensible way. 
However, in the present case, $g$ can be non-integer \cite{al-91}  and Eq. (\ref{distribution P ris}) cannot be derived with the elementary methods of 
the previous section.
Overall, the net effect of the (non-integer) entanglement spectrum  degeneracy in CFT is  the 
same found in the previous section for integer degeneracy.
In particular the number distribution in CFT is, alike Eq. (\ref{g-n}), given by
\be
n(\lambda)= g I_0(\xi_\lambda)\,.
\label{g-n2}
\ee

We mention that Eqs. (\ref{distribution P ris}) and (\ref{g-n2}) are valid also for the case of 
a finite system of length $L$ with periodic boundary conditions. 
Indeed in this case, the worldsheet for $\rho_A$ can be mapped in the one of Fig. \ref{fig1} by a conformal map. 
Thus, the only change will be the width of the annulus which becomes $W=2\ln [ (L/\pi\epsilon) \sin(\pi\ell/L) ]+ O(\epsilon)$, but this only 
affects the value of $\lambda_{\rm max}$ and not Eq.~(\ref{distribution P ris}).
Finally in \ref{sec:app_bdy} we show how the above results minimally change in the presence of  physical boundaries in CFT:
the main difference is that there are two boundary entropies which may be different, one corresponding to the boundary entropy
of the physical edge and the other being the boundary entropy introduced through the regularisation procedure at the entangling
point, as done above.

\section{Analytic results for the spin-1/2 XXZ chain}
\label{sec4}

In this section we explore the general results of the previous sections 
in some integrable spin chains, for which the entanglement spectrum can be accessed exactly. 
We present explicit results for the gapped XXZ and Ising spin-chains and exploit 
their scaling behaviour close to quantum critical points.
We use gapped spin-chains because the scaling of the moments of the reduced density matrices 
are identical to the CFT ones close to the quantum critical points and this implies that also the entanglement spectrum must be the same.
Furthermore, the entanglement spectrum  of these gapped spin-chains is easily handled analytically, as shown in the following, 
but the same is not true even for the simplest gapless lattice models.

\subsection{The anisotropic Heisenberg spin-chain.}

We consider here anisotropic Heisenberg spin-chain (XXZ spin chain), which is defined by the 
Hamiltonian\footnote{ Hereafter $\Delta$  should not be confused with the label of the 
CFT energy spectrum of the previous section}.
\be
H_{XXZ}=  \sum_{j}\left[ \sigma^x_j\sigma^x_{j+1}+ \sigma^y_j\sigma^y_{j+1}
+\Delta\sigma^z_j\sigma^z_{j+1}\right] \,,
\label{HamXXZ}
\ee
where $\sigma_i^{x,y,z}$ are the Pauli matrices and $\Delta$ is the anisotropy. 
We focus on the gapped and antiferromagnetic regime for $\Delta>1$. At $\Delta=1$ there 
is a conformal quantum critical point, separating the gapped antiferromagnetic phase from 
a gapless conformal one. For $\Delta\to1^+$, the correlation length diverges, and 
the scaling of the moments of $\rho_A$ is given by Eq.~(\ref{mom-cft}) with $\ell$ 
replaced by $\xi$ \cite{cc-04,w-06,ccp-10}. This is true in the case of a bipartition in 
two semi-infinite lines, but also for a finite interval, as long as its length 
$\ell$ is much larger than $\xi$ (for smaller $\ell$ a complicated crossover to the conformal 
results takes place~\cite{ccd-08}). 

For the case of a bipartition in two semi-infinite lines, the RDM can be written 
from the corner transfer matrix as~\cite{pkl-99,cc-04,ccp-10}
\be
\rho_{A}=\frac{e^{-H_{\rm CTM}}}{{\rm Tr}\, e^{-H_{\rm CTM}}}\,,
\ee
where $H_{\rm CTM}$ is  \cite{pkl-99} 
\be
H_{\rm CTM}=\sum_{j=0,1}^\infty \epsilon_j \hat n_j\,,
\label{hctm}
\ee
 where $\hat n_j$ are fermion number operators with eigenvalues $0$ and $1$ and
\be 
\epsilon_j= 2j\epsilon,
\ee
  with
\be  
   \epsilon={\rm arccosh} \Delta\,.
\ee
A very important point for our paper is the initial value of $j$ in the sum in~\eref{hctm}. 
Indeed, $j=0$ and $j=1$ correspond to different ground-states of the model.
If the sum starts with $j=1$, we are considering the symmetry breaking 
state (i.e. the one that for $\Delta\to\infty$ converges to the N\'eel state). 
In this state,  the largest eigenvalue of the RDM is non-degenerate. 
Conversely, if the sum starts from $j=0$, we are considering 
the combination N\'eel plus its translated by one site (usually called
anti-Neel); the latter has zero staggered magnetisation and it does not 
break the $Z_2$ symmetry  (but does not satisfy cluster decomposition, which 
is a fundamental property of physical states). 
In this case, the largest eigenvalue of the RDM is 
doubly degenerate, and the same is true for all the spectrum, as a consequence 
of the zero mode with $j=0$.

Following~\cite{pkl-99,alba-2013a,mac-17}, the entanglement spectrum is 
obtained by filling in all the possible ways the single particle levels $\epsilon_j$ 
in~\eref{hctm} (i.e., setting all $\hat n_j$ equal either to $0$ or $1$). 
The resulting  eigenvalues  of the reduced density matrix, with $n=\sum_j j$ (cf.~\eref{hctm}), are 
\be
\lambda_n= \lambda_0 e^{-2n \epsilon},
\label{lamn}
\ee
(i.e. the logarithm of the eigenvalues, usually called entanglement levels, are equally spaced with spacing $2\epsilon$).
The {\it degeneracy}  of the $n$-th eigenvalue is given by the number of ways of obtaining $n$ as a sum of smaller 
non-repeated integers. This is the problem of counting the number $q(n)$ of 
(restricted) partitions of $n$. The number $q(n)$ is conveniently generated 
as a function of $n$ via the generating function
\begin{equation}
\label{g}
G(z)\equiv\sum_{k=0}^{\infty}q(k)z^k=\prod_{k=1}^{\infty}\left(1+z^k \right). 
\end{equation}
Thus, $q(n)$ is obtained from~\eref{g} as the coefficient of the monomial $z^n$. 
The degeneracy of the entanglement spectrum is given either by $q(n)$ or by $2q(n)$ depending on whether 
we are considering the symmetry breaking state or  the symmetric one.
 This degeneracy of the entanglement  spectrum can be written in a compact way as 
\be 
{\rm deg}(\lambda_n)= \gamma q(n),
\label{degn}
\ee
 with $\gamma=1$ or $\gamma=2$ depending on the considered ground-state.
The constant $\lambda_0$ in (\ref{lamn}) is the largest eigenvalue of the RDM and can be simply
fixed by the normalisation condition 
\be\fl
\Tr \rho_A=1= \sum_n \gamma q(n) \lambda_0 e^{-2n\epsilon}= \lambda_0 \,\gamma G(e^{-2\epsilon})\, 
\qquad\Longrightarrow\qquad 
\lambda_0=\frac1{\gamma G(e^{-2\epsilon})}.
\ee

 Having obtained the analytical expression for the eigenvalues of the RDM and their degeneracies, 
it is now straightforward to write both entanglement entropies and distribution function of eigenvalues.
For example, the moments are
\be
\Tr\rho_A^n=\sum_{k=0}^\infty \gamma q(k)\lambda_0^n e^{-2nk\epsilon}=\lambda_0^n \gamma 
G(e^{-2n\epsilon})= \gamma^{1-n} \frac{G(e^{-2n\epsilon})}{(G(e^{-2\epsilon}))^n}.
\label{mome}
\ee
Notice the occurrence of the factor  $\gamma^{1-n}$, in agreement with (\ref{R-rel}).

\begin{figure}[t]
\begin{center}
\includegraphics*[width=0.85\linewidth]{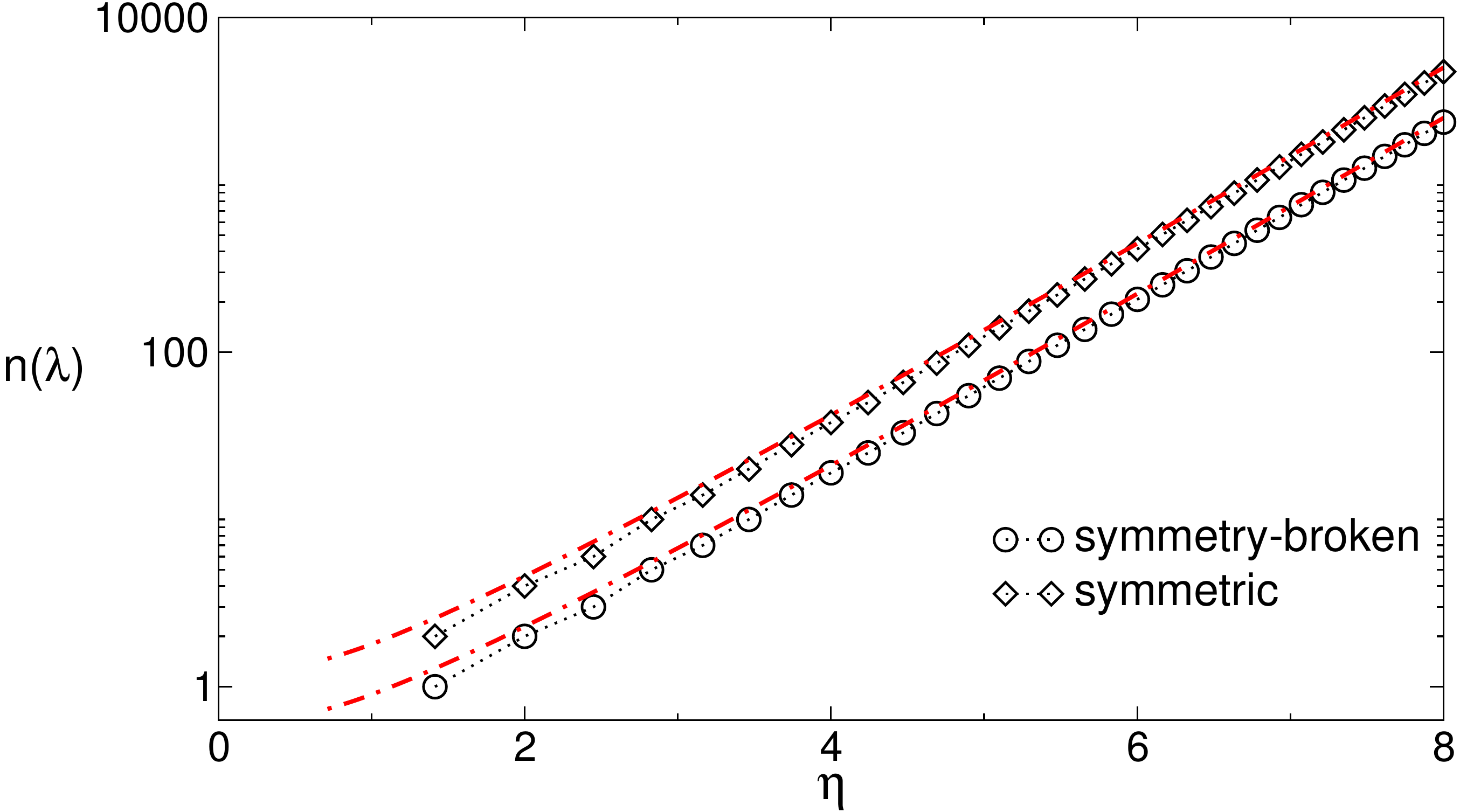}
\end{center}
\caption{ Entanglement spectrum of the XXZ chain in the gapped phase. 
The number distribution function $n(\lambda)$ is plotted versus $\eta\equiv(\ln(\lambda_0/\lambda)/\epsilon)^{1/2}$. 
The symbols are the exact  results in the thermodynamic limit  from Eq. \eref{nxxzgen} while 
the dashed-dotted lines are the results for large $\eta$ in Eq. \eref{nxxz1}. 
The two sets of data correspond to $\gamma=1$ (symmetry-broken) and $\gamma=2$ (symmetric) ground-states. 
}
\label{fig2}
\end{figure}

 Let us now move to the number distribution function $n(\lambda)$, i.e. the number of eigenvalues of the RDM 
larger than  $\lambda$. By definition this is nothing but the sum of the degeneracies ${\rm deg}(\lambda_j)$ of all eigenvalues 
$\lambda_j$ larger than $\lambda$:
\be
n(\lambda)=\sum_{\lambda_j>\lambda} {\rm deg}(\lambda_j), 
\ee
which using (\ref{lamn}) and (\ref{degn}) can be written as
\be
n(\lambda)=\sum_{m<\frac{\ln\lambda_0/\lambda}{2\epsilon}} \gamma q(m)\,,
\label{nxxzgen}
\ee
which can be straightforwardly evaluated up to very large $j$.
Here however, we are interested in the asymptotic behaviour of $n(\lambda)$
which can be obtained by expanding $q(m)$ for large $m$, 
using Meinardus theorem as in \cite{mac-17} to get
\be
\label{qn-asi}
q(m)=\frac{1}{4\cdot 3^{1/4}m^{3/4}}\exp(\pi\sqrt{m}/\sqrt{3}). 
\ee
At this point,  the asymptotic behaviour of the number distribution 
$n(\lambda)$ of the entanglement spectrum for  small $\lambda$, i.e., 
for $\ln \lambda_0/\lambda\gg \epsilon$,  can be easily calculated integrating~\eref{qn-asi} to obtain 
\be
n(\lambda)\simeq \gamma \frac{6^{1/4} e^{\pi \eta/\sqrt6}}{2\pi \sqrt{\eta}},
\label{nxxz1}
\ee
where we introduced the scaling variable
\be
\eta\equiv \sqrt{\frac{\ln \lambda_0/\lambda}{\epsilon}}\gg1.
\ee
Notice that Eq. \eref{nxxz1} is valid for arbitrary $\epsilon$ in the regime when $\eta\gg1$.
Eq.~\eref{nxxz1} is numerically  tested in Fig.~\ref{fig2}. 
The Figure shows the exact result for $n(\lambda)$  obtained using Eq. \eref{nxxzgen} (symbols in the figure) plotted versus 
$\eta$. 
The dash dotted line is~\eref{nxxz1}, and it is in perfect agreement with the exact result  for large enough $\eta$. 

As already mentioned, $\Delta=1$ corresponds to a quantum critical point with a correlation length diverging as
\be
\label{corr-l}
\ln \xi=\frac{\pi^2}{2\epsilon}+{\mathcal O}(\epsilon^0). 
\end{equation}
Thus for $\Delta\to 1^+$, Eq. (\ref{nxxz1}) should match the CFT scaling \eref{g-n2} with the appropriate value of $g$.
In order to work out this value of $g$, we can consider the R\'enyi entropies in the limit $\Delta\to 1^+$ (or $\epsilon\to 0$).
Eq. (\ref{mome}) can be expanded close to $\epsilon\to 0$ by standard methods~\cite{ccp-10}, 
obtaining
\be
S_n=\frac{\pi ^2 }{24 \epsilon}\left(1+\frac1n \right)-\frac{\ln2}2+\ln \gamma +\dots, 
\label{sn2}
\ee
where the dots stands for exponentially small terms in $\epsilon$ \cite{ccp-10}.
 We can interpret the $n$-independent additive constant (in $\epsilon$) as the non-integer degeneracy 
of the ground state of the boundary CFT with the boundary condition induced at the entangling surface. 
Thus one has
\be
g= \frac{\gamma}{\sqrt2}\,,
\label{g-xxz}
\ee
i.e. $g= 1/\sqrt{2}$ or $g=\sqrt{2}$ depending on whether $\gamma=1$ or $\gamma=2$.

We now compare the exact result~\eref{nxxz1} with the expected behaviour from conformal field theory. 
In the limit $\Delta\to1^+$ (equivalently, $\epsilon\to0$) the moments of the RDM exhibit conformal scaling. 
In this limit, taking into account the non-integer degeneracy $g$ of the ground state, the 
the number distribution function is Eq. (\ref{g-n2}) with $g$ given in \eref{g-xxz}, i.e., 
\be
n(\lambda)\simeq \gamma I_0(\xi_\lambda)/\sqrt{2}\,,
\label{nxxz2}
\ee
with $\xi_\lambda$ defined in \eref{n-lam}.
Using $-\ln \lambda_0\sim \pi^2/24\epsilon+O(1)$ (cf. (\ref{sn2})  for $n\to\infty$) 
one has $\eta\simeq \sqrt{6} \xi_\lambda/\pi$ for $\epsilon\to0$. 
We then have that,  both for $\gamma=1$ and $\gamma=2$, Eq. (\ref{nxxz1}) in the limit $\epsilon\to 1$ coincides with 
Eq. (\ref{nxxz2}) for large $\xi$, given that $I_0(x)\simeq e^{x}/\sqrt{2\pi x}$ for $x\gg1$. 
 Notice that for the XXZ spin-chain in the regime considered here, the entanglement spectrum distribution function  
of \cite{cl-08} (i.e. \eref{g-n2} with $g=1$) does not correctly describe the spectrum for none of the two possible ground states.
 
\subsection{The Ising spin-chain in a transverse field}

 An analysis similar to the one presented in the previous section holds also for the transverse field Ising chain, 
which is defined by the Hamiltonian 
\be
H=-\sum\limits_{j=1}^L(\sigma_j^x\sigma_{j+1}^x+h\sigma_j^z),  
\ee
where $h$ is the external transverse magnetic field. 
At zero temperature the model is paramagnetic for $h>1$, whereas it is an ordered ferromagnet for $h<1$. 
At  $h=1$ there is a quantum critical point with a low-energy spectrum described by a CFT with $c=1/2$. 
In the ferromagnetic region, in the limit $L\to\infty$, there are two degenerate ground states, which are related by a $\mathbb{Z}_2$ 
symmetry.  
In contrast, in the paramagnetic region $h>1$ the ground  state is unique. 

The entanglement spectrum in both regimes has the same structure~\eref{hctm} as for the XXZ chain, 
although now the single particle entanglement spectrum levels are given by \cite{pkl-99}
\be
\label{ises}
\epsilon_j=\left\{
\begin{array}{cc}
(2j+1)\epsilon & h>1,\\
2j\epsilon     & h<1,
\end{array}
\right.
\ee
with
\be
\label{epsis}
\epsilon\equiv\pi\frac{K(\sqrt{1-k^2})}{K(k)},\qquad k\equiv\textrm{min}(h,h^{-1}). 
\ee
with $K(x)$ the complete elliptic integral of the first kind. 

 It should be clear that, due to the form of the entanglement levels \eref{ises},
in the symmetry broken phase for $h<1$ the entanglement spectrum distribution  can be 
derived identically to the previous section.
Once again, the distribution depends on whether one considers the symmetric broken or symmetric ground-state,
obtained by letting the sum in $j$ starting either from $0$ or $1$.
The resulting distribution is then \eref{nxxz2} with $\gamma$ given by \eref{g-xxz}.

 In the paramagnetic phase for $h>1$ the entanglement spectrum distribution is instead different from the 
one of the XXZ model. In this case,  the degeneracy of the entanglement spectrum at level $m$ is related 
from~\eref{ises} to the number of integer partitions $q_o(m)$ of $m$ involving only {\it odd} integers
 (in this case the sum always starts from $j=1$ because the ground-state is unique). 
The asymptotic behavior for large $m$ of $q_o(m)$ is given as~\footnote{The sequence $q_o(m)$ is reported as 
A000700 in the \href{https://oeis.org}{OEIS}.} 
\be
q_o(m)\simeq\frac{1}{2\cdot 24^{1/4}m^{3/4}}\exp(\pi\sqrt{m}/\sqrt{6}),
\ee
and the corresponding distribution $n(\lambda)$ is 
\be
\label{nis-2}
n(\lambda)\simeq \frac{6^{1/4} e^{\pi \eta/\sqrt6}}{\sqrt{2\eta}\pi }\,,
\ee
 where, as usual,  $\eta\equiv \sqrt{{\ln(\lambda_0/\lambda)}/{\epsilon}}\gg1$.
Approaching the critical point at $h=1$, the correlation length diverges as 
\be
\label{corr-is}
\ln\xi=\frac{\pi^2}{\epsilon}+{\mathcal O}(\epsilon^0),
\ee
and the R\'enyi entanglement entropy as \cite{ccp-10}
\be
S_n= \frac{\pi^2}{24 \epsilon} \left(1+\frac1n\right) +O(\epsilon)\,.
\label{S-n-is}
\ee
Notice in particular the absence of the constant term, suggesting $\ln g=0$.
By replacing $\epsilon$ with $\ln\lambda_0$ (using~\eref{S-n-is}), one can straightforwardly show that ~\eref{nis-2} 
coincides with the large $\xi_\lambda$ limit of $I_0(\xi_\lambda)$, i.e.~\eref{g-n2} with $g=1$.  

\section{Consequences for the negativity spectrum}
\label{sec5}

The partial transpose $\rho_A^{T_2}$ of the reduced density matrix is a crucial object 
for quantifying the bipartite entanglement in a mixed states~\cite{peres-1996,zycz-1998,
vidal-2002,plenio-2005} or, equivalently, the entanglement between two non-complementary parts in a pure state. 
The partial transpose $\rho_A^{T_2}$ is defined as $\langle\varphi_1\varphi_2|\rho^{T_2}_A|\varphi'_1
\varphi'_2\rangle\equiv\langle\varphi_1\varphi_2'|\rho_A|\varphi'_1\varphi_2\rangle$, 
with $\{\varphi_1\}$ and $\{\varphi_2\}$ two bases for $A_1$ and $A_2$, respectively.
A computable measure of entanglement is the {\it logarithmic negativity} defined as 
the sum of the absolute values of the eigenvalues of $\rho_A^{T_2}$ \cite{vidal-2002,plenio-2005}
\begin{equation}
\label{neg}
{\cal E}\equiv \ln||\rho_A^{T_2}||_1=\ln\textrm{Tr}|\rho^{T_2}_A|,
\end{equation}
where the symbol $||\cdot||_1$ denotes the trace norm. 
The scaling behaviour of the negativity has been characterised analytically for the 
ground states of one dimensional CFTs~\cite{calabrese-2012,cct-neg-long,calabrese-2013}. 
Remarkably, the negativity is scale invariant at  quantum critical points~\cite{hannu-2008,mrpr-09,
hannu-2010,calabrese-2012}. Its scaling behaviour has been also worked out for finite 
temperature CFTs~\cite{calabrese-2015}, in CFTs with large central charge~\cite{kpp-14}, 
in disordered spin chains~\cite{ruggiero-2016}, in some holographic~\cite{rr-15} and 
massive quantum field theories~\cite{fournier-2015}, for out of equilibrium 
models~\cite{ctc-14,hoogeveen-2015,eisler-2014,wen-2015}, topologically ordered 
phases~\cite{lee-2013,castelnovo-2013}, Kondo-like systems~\cite{bayat-2012,bayat-2014,
abab-16}, and Chern-Simons theories~\cite{wen-2016,wen-2016a}. 
Surprisingly, no analytical results are available yet for free-fermion models~\cite{neg-ff}, 
in contrast with free bosonic model, for which the negativity can be 
calculated~\cite{audenaert-2002}, also in $d>1$ dimensions~\cite{eisler-2016, dct-16}. 

On the same lines as for the entanglement spectrum, it is clear that the partial transpose 
contains more information than that condensed in the negativity. In analogy with what 
explained above, part of this information can be reconstructed from the scaling of the 
moments of the reduced density matrix in a program initiated in~\cite{rac-16}. 
As a fundamental difference compared to the entanglement spectrum, 
the partial transpose has both positive and negative eigenvalues which generically behave in a different manner.
However, their asymptotic CFT distribution for very small eigenvalues turned out to be the same \cite{rac-16}.

Here we  first focus on the case of a bipartition of the ground state of a CFT, when the spectrum 
of the partial transpose can be written in terms of the spectrum of the RDM. 
Although in this case the partially transposed density matrix does not contain more 
information than the reduced density matrix itself (the two spectra can be simply 
related~\cite{rac-16}), it is very useful to obtain the distribution of these 
eigenvalues to understand how  the CFT non-integer ground-state degeneracy induced at the entangling surface
affects the results for the negativity. 
For this bipartition, the moments of $\rho^{T_2}$ can be put in direct relation with 
those of $\rho_{A_2}$ as shown in~\cite{calabrese-2012}, obtaining  
\be
\Tr (\rho^{T_2})^n =
\left\{
\begin{array}{ll}
\textrm{Tr} \rho_{A_1}^{n_o} =  c^T_{n_o} \ell^{-c/6(n_o-1/n_o)} & n_o\;\textrm{odd},\\
(\textrm{Tr} \rho_{A_1}^{n_e/2})^2=c^T_{n_e} \ell^{-c/3 (n_e/2-2/n_e) }& n_e\;\textrm{even}.
\end{array}
\right.
\ee
The relation between $c^T_n$ and $c_n$ depends on the parity of $n$ and it reads as 
\bea
c^T_{n_o}&=&c_{n_o}, \nonumber\\
c^T_{n_e}&=&c_{n_e/2}^2. 
\label{ctc}
\eea
In Ref. \cite{rac-16} the negativity spectrum distribution in CFT has been derived with the assumption 
$c_n=1$ that implies $c^T_n=1$. 

It is now easy to understand the effect of a degeneracy (both integer and non-integer).
We have shown that in the presence of degeneracy $g$, cf.~Eq.~(\ref{c-lam-mu}), the multiplicative 
constant $c_n$ gets multiplied by $g^{1-n}$, implying a non-trivial relation between the 
constants $c_n^T$ for the moments of the partial transpose. 
According to the equations in~(\ref{ctc}) we have 
\bea
c_{n_o}^T(\lambda)=g^{1-n_o}, \nonumber\\
c_{n_e}^T(\lambda)=(g^{1-n_e/2})^2= g^{2-n_e}\,.
\eea
In \cite{rac-16} it was shown that the number distribution function for the case $c_n=1$ is 
\begin{equation}
\label{pure-n}
n^{\rm pure}(\lambda)= \frac{1}{2}\big[\textrm{sgn}(\lambda)I_0(\xi_\lambda)+I_0(2\xi_\lambda)\big],
\end{equation}
where $\xi_\lambda$  is defined in analogy with Eq. (\ref{n-lam}) as 
\be
\xi_\lambda=2\sqrt{(-\ln \lambda_{\rm Tmax})\ln(\lambda_{\rm Tmax}/|\lambda|)}\,,
\label{xi-T}
\ee
where $ \lambda_{\rm Tmax}$ is the maximum positive eigenvalue of the partial transpose
(which in the case of a bipartite pure system is equal to $\lambda_{\rm max}$ \cite{rac-16}, but not in general).
It is easy to check, following the derivation in~\cite{rac-16}, that the first term comes 
from the odd moments of the partial transpose, while the second from the even ones. 

Thus, to take into account the degeneracy $g$ in~\eref{pure-n}, the first term gets 
a factor $g$, whereas the second a factor $g^2$. Consequently, in the presence of a 
global degeneracy $g$, the negativity spectrum distribution is 
\begin{equation}
n^{\rm pure}(\lambda)=\frac{1}{2}\big[ \textrm{sgn}(\lambda)  g I_0(\xi_\lambda)+g^2 I_0(2\xi_\lambda)\big]. 
\label{n-neg-deg}
\end{equation}
This formula is quite interesting because $g$ enters in a different way in the two terms. 
 In particular it shows that for asymptotic small eigenvalues $\lambda$, i.e. large $\xi_\lambda$,
positive and negative eigenvalues have the same distribution which gets multiplied by $g^2$, but the difference of the number 
distributions (i.e. $n^{\rm pure}(\lambda)-n^{\rm pure}(-\lambda)$) gets multiplied  by $g$.

Let us briefly mention what we know about the very interesting case of a tripartite system, when the spectrum of the partially transposed 
density matrix cannot be written in term of that of the reduced density matrix, because the two intervals are in a mixed state. 
We focus on the tripartition where two finite intervals $A_1$ and $A_2$ are adjacent and embedded either in the infinite line
or in a finite system.
We consider the reduced density matrix $\rho_A=\rho_{A_1 \cup A_2}$ and subsequently the partial transpose with respect to $A_2$.
In this case, the conformal moments of  $\rho_A^{T_2}$ depend on the constants $c_n$ appearing in the moments of $\rho_{A_2}$ 
but also on some $n$-dependent structure constants \cite{calabrese-2012,cct-neg-long}. 
Assuming that all these parameters are equal to $1$, the negativity spectrum distribution has been obtained in~\cite{rac-16}
and tested against numerical simulations in spin-chains.
In the case when the two intervals have equal length, the final result can be written as \cite{rac-16}  
\begin{equation}
\label{mixed-n}
n^{\rm mixed}(\lambda)= \frac{1}{2}\big[\textrm{sgn}(\lambda)I_0(\xi_\lambda)+I_0(\sqrt{2}\xi_\lambda)\big],
\end{equation}
 where $\xi_\lambda$ is defined in Eq. \eref{xi-T} and in this case $\lambda_{\rm Tmax}$ is different from $\lambda_{\rm max}$. 
Note the $\sqrt{2}$ difference in the argument of $I_0$ as compared with~\eref{pure-n}. 
 At this point, while it is known how the degeneracy factor $g$ affects the multiplicative factors $c_n^T$, 
the same is not true for the structure constants  and its effect is not trivial. 
 Indeed, for $n=2$ we have ${\rm Tr} (\rho^{T_2}_A)^2= {\rm Tr} \rho_A^2$ \cite{cct-neg-long}
implying that these structure constants depend on $g$ and we cannot simply set to a constant when $g\neq 1$.
In \cite{mac-17}, the moments of the partial transpose, as well as the entire negativity spectrum have been 
analytically worked out for the XXZ spin-chain in the gapped regime discussed in the previous section. 
While the leading behaviour for large correlation length $\xi$ is the same as in CFT for large $\ell$ (providing to two different exponentials  
in $n(\lambda)$ as in (\ref{mixed-n})), the subleading terms are different and so not useful to understand what happens for $g\neq 1$ in a CFT. 
It would be then very interesting to run some numerical simulations in gapless models with $g\neq1$ to shed some light on this problem.

\section{Conclusions}
\label{sec6}

In this paper we established a relation between the entanglement spectrum 
distribution of conformal field theories (see Ref.~\cite{cl-08}) and the CFT 
density of states described by the Cardy formula~\cite{c-86}. Indeed, we have 
shown that the entanglement spectrum distribution can be obtained as a 
re-parametrisation of the Cardy formula. This result allows us to 
understand the effect of the boundary conditions at the entangling surface 
on the entanglement spectrum and entropies. 
In particular, it shows that the multiplicative constant $c_n$ in the moments 
of the reduced density matrix, cf. Eq. (\ref{mom-cft}), is affected by the 
degeneracies induced by these boundary conditions, although the leading behaviour 
in $\ell$ of the moments remains the same. We tested our findings against exact 
calculations of the entanglement spectrum in integrable spin chains. 
 Furthermore our result shows that the Affleck-Ludwig entropy due to the boundary conditions 
induced at the entangling surface can be measured  from the asymptotic behaviour of the entanglement  
spectrum distribution. 
This proposal appears to be more practical than the study of the additive constant of the R\'enyi entropies 
whose value in microscopical models is generically influenced by the non-universal ultraviolet physics \cite{jk-04,ij-08,ce-10,cmv-11,dsvc17}.

We also explored the consequences of degeneracies for the negativity spectrum. 
In contrast with the entanglement spectrum distribution, 
which is expressed in terms of a single Bessel function, the negativity 
spectrum is the combination of two Bessel functions. 
Remarkably, the presence of a global degeneracy in the entanglement spectrum gives rise to different 
 reparametrisation of the two functions. 

%
A consequence of our findings is that the several numerical results 
already present in the literature \cite{cl-08,pm-10,ahl-12,lr-14,si-14,
laflorencie-2016} about the validity of the CFT results for the entanglement 
spectrum distribution (\ref{n-lam}) can be regarded as direct verifications of 
the Cardy formula, which,  instead, has never been checked at the level of the 
hamiltonian spectrum of microscopic models. 
The reason  why Cardy formula has not been tested from the hamiltonian spectrum 
is that CFT describes the low-energy spectrum of microscopic models where the dispersion 
relation is relativistic, while Cardy formula gives the scaling of the 
CFT spectrum for large energy, where in microscopic models non-relativistic effects become relevant.
Conversely, the entanglement spectrum  distribution is  written only in terms of the ground-state, and a continuum distribution 
for the eigenvalues is obtained in the limit $\ell\to\infty$.

\section*{Acknowledgments}
VA thanks the funding from the European Union's Horizon 2020 research and 
innovation programme under the Marie Sklodowoska-Curie grant agreement No 702612 OEMBS.

\appendix

\section{ The Cardy formula on the annulus}
\label{sec:app}

In this section we report a derivation of the Cardy formula starting from the partition function of a CFT on the annulus. 
 A similar analysis has been done by employing the geometry of the torus in \cite{loran,carlip-00} 
 and of the Klein bottle in \cite{tu-17}.
The aim of this computation is to show the occurrence of the boundary states of the annulus 
in the multiplicative factor of the CFT density of states for large scaling dimensions. 
This factor corresponds to the ground state degeneracy introduced by Affleck and Ludwig~\cite{al-91}.

The starting point is the partition function on the annulus of width 
$W$ \cite{c-89} written as a sum over the states parametrised by their conformal 
dimension $\Delta$
\be
\mathcal{Z}(q) = q^{-c/24} \sum_\Delta \rho(\Delta)\, q^\Delta,
\label{Zq}
\ee
where $q=e^{-2\pi^2/W}$ is the modular parameter. Here $\rho(\Delta)$ is the 
Cardy density of states which can be formally obtained by inverting the above 
relation as a complex integral in $q$
\be
\rho(\Delta) = \frac{1}{2\pi \iu}
\oint_{C_0} \frac{dq}{q^{\Delta + 1}}\, q^{c/24} \mathcal{Z}(q),
\label{rho-int}
\ee
in which $C_0$ is an arbitrary closed path  which encloses $q=0$. 

By employing the modular invariance, the partition function (\ref{Zq}) 
can be written in terms of the dual modular parameter $\tilde q=e^{-2W}$ as \cite{c-86}
\be
\mathcal{Z}(q) = \tilde{q}^{-c/24}  \sum_k \langle a | k \rangle  \langle k | b \rangle \,\tilde{q}^{\delta_k},
\ee
where the sum is now over all allowed scalar bulk operators with 
dimensions $\delta_k$, and $|a\rangle$ and $|b\rangle$ denote the boundary 
states at the two edges of the annulus. Plugging this into (\ref{rho-int}) we have  
\be
\rho(\Delta) 
= \frac{1}{2\pi \iu}
\oint_{C_0} \frac{dq}{q^{\Delta + 1}}\, q^{c/24} \, \tilde{q}^{-c/24}  \sum_k \langle a | k \rangle  \langle k | b \rangle \,\tilde{q}^{\delta_k}.
\ee
Let us know change the integration variable as 
\be
q=e^{-t},
\qquad
\tilde{q} = e^{-4\pi^2/t}.
\ee
Choosing $C_0$ to be a circumference centred at the origin $q=0$, we have
\bea
\rho(\Delta) 
&=&
\sum_k 
\langle a | k \rangle  \langle k | b \rangle  \,
\frac{1}{2\pi \iu}
\int^{t_1 + \iu \pi}_{t_1 - \iu \pi}
dt\, e^{(\Delta - c/24) t}\, e^{(\gamma-\gamma_k)/t},
\eea
where we introduced 
\be
\gamma \equiv  \frac{\pi^2  c}{6},
\qquad
\gamma_k \equiv  4\pi^2 \delta_k.
\ee
Hereafter $\gamma$ must not be confused with the integer degeneracy factor in the main text.

The above equation is an exact representation of the CFT density of states, valid for arbitrary values of $\Delta$, but it 
is particularly convenient for a large $\Delta$ expansion. 
Indeed, for large $\Delta$ two main simplifications occurs. 
First, the integrals are dominated by the saddle points at $t=0$, thus one can extend the integrals on the imaginary axis 
between $\pm \iu \infty$ instead of $\pm \iu \pi$.
In the same spirit, it is reasonable to keep only the term with $k=0$, being the others expected to be exponentially suppressed.
Thus one has
\bea
\label{inv-lap}
\rho(\Delta)  
&\simeq&
\langle a | 0 \rangle  \langle 0 | b \rangle  \,
\frac{1}{2\pi \iu} 
\int^{t_1 + \iu \infty}_{t_1 - \iu \infty}
dt\, e^{(\Delta - c/24) t}\, e^{\gamma/t}
\\
&=&
\langle a | 0 \rangle  \langle 0 | b \rangle  
\left[
\delta(\Delta - c/24) + 2\gamma  \frac{I_1\big(2\sqrt{\gamma(\Delta - c/24)}\,\big)}{2\sqrt{\gamma(\Delta - c/24)}}
\right] 
\nonumber \\
&=&
\langle a | 0 \rangle  \langle 0 | b \rangle  
\left[
\delta(\Delta - c/24) + \frac{\pi^2 c}{3}\;\frac{I_1\big(2\pi\sqrt{(c/6)(\Delta - c/24)}\,\big)}{2\pi\sqrt{(c/6)(\Delta - c/24)}}
\right], \nonumber
\eea
where $I_1$ is modified Bessel function of the first kind that is obtained by the inverse Laplace transform in (\ref{inv-lap}).

This result is employed in Sec.\,\ref{sec3} in a crucial way to recover  for $g=1$ the result of \cite{cl-08}
from the CFT entanglement spectrum found in \cite{ct-16}.
The formula (\ref{inv-lap}) exposes the role of the one-point structure constants $\langle 0 | a \rangle$ and $ \langle 0 | b \rangle$ corresponding to the boundary 
states which characterise the underlying CFT on the annulus.
The term $\delta(\Delta - c/24)$ is often neglected in the literature (see however \cite{behan}), 
because it is subleading for large $W$, but we showed it in order to stress the equivalence with the entanglement spectrum \cite{cl-08}. 
It is fair to mention that a very similar analysis, but in a slightly different context, has been presented also in \cite{rajap}.

We conclude this appendix by reporting the expansion of the density of states (\ref{inv-lap}) for large $\Delta$.
When $\Delta \gg 1$, we can use the following asymptotic behaviour of the Bessel functions 
\be
I_1(z) = \frac{e^z}{\sqrt{2\pi z}} \left( 1 -\frac{3}{8\,z}+ O(1/z^2)\right),
\ee
to expand (\ref{inv-lap}) as
\be 
\label{rho AL order1}
\fl
\rho(\Delta) =
\langle a | 0 \rangle  \langle 0 | b \rangle  
\,\frac{\sqrt{\gamma}}{\sqrt{\Delta}}
\, \frac{e^{2\sqrt{\gamma\Delta}}}{\sqrt{4\pi(\gamma\Delta)^{1/2}}}
=
\langle a | 0 \rangle  \langle 0 | b \rangle  
\left(\frac{c}{96 \Delta^3}\right)^{1/4} e^{2\pi \sqrt{c\Delta/6}},
\ee
which is the result found in \cite{al-91}.

\section{Entanglement spectrum distribution in  the presence of boundaries}
\label{sec:app_bdy}

\begin{figure}[t]
\begin{center}
\includegraphics*[width=0.65\linewidth]{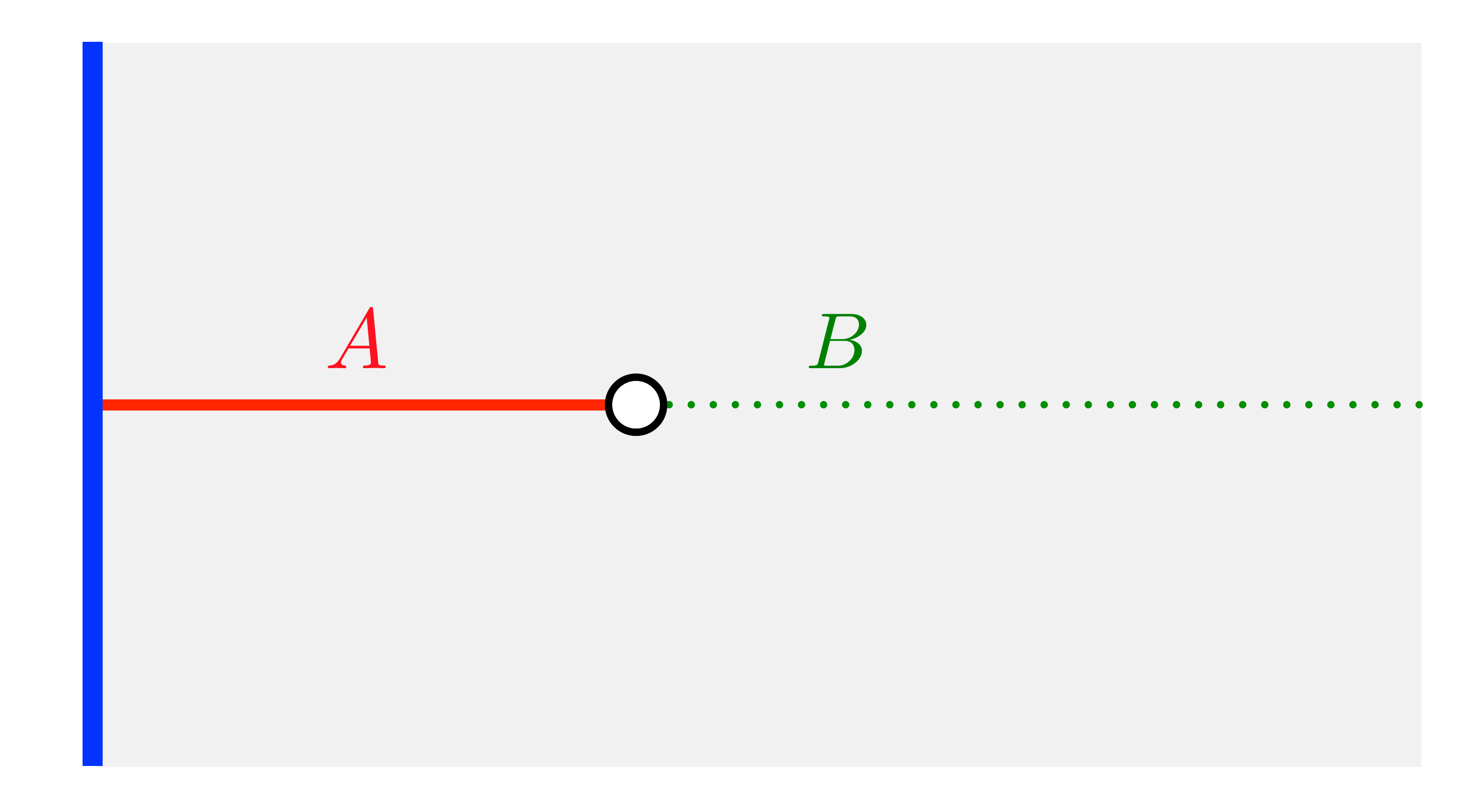}
\end{center}
\caption{ Path integral for reduced density matrix $\rho_A$ of an interval of 
 length $\ell$ at the edge of a semi-infinite system whose boundary in the Euclidean spacetime
 is the solid blue line.
 The field theory is regularised by removing a small disk of radius $\epsilon$ centred 
 at the point (entangling surface) which separates the interval $A$ from its complement $B$.
 Conformally invariant boundary conditions which are not necessarily the same 
 are imposed both on the physical boundary (blue solid line) and on the boundary of the small disk (black solid line).
 The moments ${\rm Tr} \rho_A^n$ are computed by joining cyclically, along $A$,  $n$ replicas of $\rho_A$. 
 The resulting manifold is an annulus which can have different conformally invariant boundary conditions at 
 its two boundaries. 
}
\label{fig_app_bdy}
\end{figure}

In this appendix we discuss how the result for the entanglement spectrum distribution slightly changes in the presence of real boundaries.
We focus on the case in which the system is the semi-infinite line $x\geqslant 0$
and the subsystem is the interval $A=[0,\ell]$. 
The corresponding Euclidean spacetime is the half space $(x,y) \in \mathbb{R}^2$ with $x\geqslant 0$,
whose boundary is the infinite line $x=0$, which supports a conformally invariant boundary condition of the model. 
Following the same regularisation procedure adopted in Sec.\,\,\ref{sec3}, we remove 
a small disk of radius $\epsilon$ centred at the entangling point $x=\ell$.
A conformally invariant boundary condition  is induced along the boundary of this small disk,
which may be different from the one along the physical boundary at $x=0$.
The resulting spacetime is depicted in Fig.\,\ref{fig_app_bdy}.
It has the topology of an annulus where different conformally invariant boundary conditions 
may be  present on the physical boundary (blue solid line) and on the boundary of the regularising disk 
(black solid line).
In \cite{ct-16} the latter observation has been employed to study $K_A$ for this configuration,
finding that for the corresponding entanglement spectrum Eq.\,(\ref{kappa}) holds, but
with $W=\log (2\ell/\epsilon) + O(\epsilon)$ in this case.

Thus,  the analysis of Sec.\,\ref{sec3} can be repeated with the crucial difference that
we cannot impose $g_a = g_b$ because we are allowed to have different conformally invariant boundary conditions. 
Denoting by $g_0$ the degeneracy introduced by the boundary encircling the endpoint $x=\ell$ and by $\tilde{g} $ 
the degeneracy corresponding to the conformally invariant boundary condition on the physical boundary, for the distribution of eigenvalues we obtain
\be\fl
\label{P_bdy}
\widetilde{P}(\lambda) = 
 \theta(\lambda_{\rm max} - \lambda)   \tilde{g} \,g_0 \,
 \frac{b}{ \lambda }\,
\frac{I_1\big( 2 \sqrt{b \ln(\lambda_{\rm max}   / \lambda ) }  \, \big)}{
\sqrt{b \ln(\lambda_{\rm max}   / \lambda) }}
+
\tilde{g} \,g_0  \, \delta(\lambda_{\rm max} - \lambda)
\,,
\ee
being $b=-\ln\lambda_{\rm max}=cW/12 = (c/12) \ln (2\ell/\epsilon) + O(\epsilon)$.
Integrating (\ref{P_bdy}), the mean number of eigenvalues larger than a given $\lambda$ is simply obtained as 
\be
\label{n_bdy}
\tilde{n}(\lambda) = \tilde{g} g_0  I_0(\xi_\lambda)\,,
\ee
where $\xi_\lambda$ is defined in terms of $\lambda_{\rm max}$ in (\ref{n-lam}).

Eq. (\ref{P_bdy}) is valid also for a finite system of length $L$ with conformal invariant boundary condition at the two ends 
and with $A=[0,\ell]$. Indeed in this case, the worldsheet for $\rho_A$ can be mapped in the one of Fig. \ref{fig_app_bdy} 
by a conformal map. 
Thus, the only change is the width of the annulus which becomes $W=\ln [ (2L/\pi\epsilon) \sin(\pi\ell/L) ]+ O(\epsilon)$, but this only 
affects the value of $\lambda_{\rm max}$ and not Eq.~(\ref{P_bdy}).

\end{document}